# A Unified Description of Superconductivity: Electron Density Dynamic's Response to Phonons at the Brillouin Zone Boundary Driven Real Space Electron-Hole Pairing Characterized by CP Symmetry

T. Guerfi[1]

Department of Physics. Faculty of Sciences. M'Hamed Bougara University, Boumerdes, Algeria.

**Abstract**

It is argued that electron density dynamic's response to longitudinal acoustical phonon modes at the Brillouin zone boundary is the key factor in the superconducting mechanism in MgB$_2$ compound as well as in all superconducting compounds leading to a real space electron-hole pairing characterized by CP symmetry (Charge and Parity symmetries) that enables a unified treatment of superconductivity phenomenon. First principle calculation of the electron density difference is determined by subtracting the electron density of the crystal distorted by longitudinal acoustical frozen-phonon mode at the Brillouin zone boundary from that of the perfect crystal. The origin of this electron density response to this kind of distortion, that leads to a real space electron-hole pairing characterized by CP symmetry, is discussed in terms of local polarizabilities and delocalized transfer of charge caused by phonon-distortion driven dynamical change of orbital hybridizations.

This approach also provides clear and quick guidelines in the search for new superconductors.

## 1. Introduction

Density Functional Theory (DFT) is a computational quantum mechanical method [1-3] that provides an effective alternative to a solution of a many-body system Schrodinger wave equation to investigate the ground state electronic properties of a crystalline solids. In DFT the electron density $n(r)$ is the basic variable in real, three-dimensional coordinate space that all properties of the system can be considered to be unique functionals of the ground state density. In this spirit

---

[1]) t.guerfi@univ-boumerdes.dz
Tel: +213555237358

DFT enables simplistic resolution of fine-scale structural variations for a set of particular crystallography [1]. The frozen-phonon method is a direct approach in which a distorted crystal is treated as a crystal in a new structure with a lower symmetry than the undistorted crystal. The ab initio evaluation of the total energy $E_{tot}$ of the solid with frozen-in atomic displacements pattern for a given phonon mode is an efficient method to study electron-phonon coupling in crystalline solids including superconducting materials [4-6].

In this manuscript, it is argued that electron density dynamic's response to longitudinal acoustical phonon modes at the Brillouin zone boundary is the key factor in the superconducting mechanism in $MgB_2$ compound as well as in all superconducting compounds leading to a real space electron-hole pairing characterized by CP symmetry (Charge and Parity symmetries) that enables a unified treatment of superconductivity phenomenon. First principle calculation of the electron density difference is determined by subtracting the electron density of the crystal distorted by longitudinal acoustical frozen-phonon mode at the Brillouin zone boundary from that of the perfect crystal. The electron density response to this kind of distortion leads to a real space electron-hole pairing characterized by CP symmetry and its origin is discussed in terms of local polarizabilities and delocalized transfer of charge caused by phonon-distortion driven dynamical change of orbital hybridizations.

This manuscript also proposes clear guidelines in the search for new superconductors.

## 2. MgB₂ superconductor

$MgB_2$ is a superconductor at comparatively high $T_c$ of 39K [7]. The crystal structure of $MgB_2$ is known as hexagonal (AlB₂ type, space group P6/mmm) with experimental lattice constants of a= 3.083 Å and c = 3.521 Å [8]. The characteristic boron honeycomb sheets are sandwiched between the Mg triangular sheets like an intercalated graphite. Fig. 1 illustrates the atomic displacement patterns for $A_{2u}$ mode and $E_{2g}$ mode at the zone-boundary A point of the hexagonal Brillouin zone. The $E_{2g}$ mode involves only in-plane displacements of B atoms. However, the acoustical $A_{2u}$ vibration mode involves the displacement of Mg atoms along the c axis as a result a compressed and elongated unit cells will alternate all along the c axis. So, a supercell obtained by doubling the elementary cell along the [001] is needed for DFT calculation. First principle calculations of the electronic band structure as well as the electron density are carried out by using the pseudopotential method. As for the exchange and correlation terms, the Perdew Burke and Ernzerhof (PBE) for solids is used within the generalized gradient approximation (GGA) [9, 10] as implemented in the Quantum Espresso package [11, 12] and Burai software [13]. Ultra-soft pseudo-potentials [14] are used to deal with interaction between the ion cores and valence electrons.

### 2.1 Electronic bands and partial DOS calculation

First principle calculation of the electronic bands of supercell (obtained by doubling the elementary cell along the c axis) of $MgB_2$ along high symmetry points in the first Brillouin zone is performed for the perfect crystal and that of the crystal distorted by longitudinal acoustical frozen-phonon modes $A_{2u}$ mode along with the $E_{2g}$ mode. The results of electronic bands structure are displayed in figs. 2 a and b. One of the most interesting features of the calculated energy band structure for the perfect crystal case (fig. 2 a) is the two doubly degenerate bands, almost flat, present just above the Fermi energy $E_F$ in the $\Gamma - A$ direction. These bands are incompletely filled $\sigma$ bands indicating that the transport properties are dominated by the holes in the plane where the B atoms exist. The hole doping of the covalent $\sigma$ bands, is reached through the ionic layered character of $MgB_2$, as has been revealed by An and Pickett [5]. The role of the intercalant Mg atom is not simply donating electrons to boron's bands. The main change upon intercalation is the downward shift of the $\pi$ bands compared to the covalent $\sigma$ bands. The attractive potential of $Mg^{2+}$ between $B_2$ layers lowers the $\pi$ bands, resulting in $\sigma \to \pi$ electron transfer that drives the hole doping of the $\sigma$ bands [5].

The electronic band structure for the crystal distorted by $E_{2g} + A_{2u}$ phonons modes at A point of the Brillouin zone boundary, is illustrated in fig2. b. One can see a splitting of the $\sigma$ bands that correspond the compressed cell from that of the elongated cell. While the $\sigma$ bands of the compressed cell moves down the Fermi level indicating a Metal-Insulator (MI) transition, $\sigma$ bands of the elongated cell move upward indicating an increase of the hole doping level. The MI transition of the compressed cell is evidenced by the shift in energy of the electronic states of B(1) and B(2) down the Fermi level as can be seen in fig. 3 which represents the partial DOS of the $p$ orbitals of B(1), B(2) and that of B(3) and B(4). This MI transition in the compressed cell can be understood in terms of local polarizability and charge transfer caused by phonons-distortion driven dynamical change of orbital hybridizations. The valence band in $MgB_2$ is mainly composed of B–$2s$, $2p$ states hybridized with small amount of Mg–$2p$, $3s$. When the unit cell is compressed, the Mg-B(1, 2) bond along the c axis becomes short, the $p_z - s$ hybridization increases and the charge (electron) is partially transferred from B(1, 2) to Mg creating a current. In the elongated unit cell, the Mg-B(3, 4) bond along the c axis becomes elongated, the hybridization becomes less and electrons, partially move back from Mg to B(3, 4). In fact, the $E_{2g}$ mode plays also a crucial role in the Mg-B distance along the c axis and therefore in the dynamical change of orbital hybridizations. Accordingly, the attractive potential of $Mg^{2+}$ between $B_2$ layers in the compressed cell will rise the energy of the $\pi$ bands resulting in $\pi \to \sigma$ electron transfer therefore an MI transition , however it is exactly the opposite that happen in the elongated cell; the attractive potential of $Mg^{2+}$ between $B_2$ layers will lower more the energy of the $\pi$ bands relative to $\sigma$ bands , resulting in further $\sigma \to \pi$ electron transfer that drives more hole doping of the $\sigma$ bands. One also notes that these Metal-Insulator and Insulator-Metal transitions will oscillate with the $A_{2u}$ frequency.

It should be pointed out that a polarization catastrophe is evidenced to occur for $YBa_2Cu_3O_{7-\delta}$ system undergoing at the superconducting critical temperature a Metal-Superconductor (MS) transition [15, 16] . A model of Metal-Insulator MI transition that was proposed by Goldhammer and Herzfeld [17] was the polarization catastrophe. The criterion is based on the Clausius-Mossotti relation of classical electrostatics. This relates the atomic polarizability $\alpha$, the atomic volume $v$ and relative permittivity $\varepsilon$

$$\varepsilon = \frac{2\left(\frac{4\pi\alpha}{3v}\right) + 1}{1 - \left(\frac{4\pi\alpha}{3v}\right)}$$

If the atomic volume is decreased by decreasing the temperature (or by applying a pressure), at given critical temperature $T_c$ we can reach a situation in which

$$\frac{4\pi\alpha}{3v} \to 1$$

there is a polarization catastrophe and $\varepsilon \to \infty$.

Which suggests therefore the same criterion for the Metal-Superconductor MS transition.

## 2.2 Electronic density and electronic density difference calculation

First principle calculation of the Electron Density Difference (EDD) is determined by subtracting the electron density of the crystal distorted by frozen-phonon $A_{2u}$ and $E_{2g}$ modes at the Brillouin zone boundary from that of the perfect crystal. The electron density in the perfect crystal and that of the distorted crystal are depicted in Figs. 4 a and b respectively, using VESTA software [18]. A comparison between them reveals that there is more electronic polarizability in the compressed cell and less polarizability in the elongated cell relative to the electron density of the perfect crystal. Subtracting the electron density of the crystal distorted from that of the perfect crystal, we get the EDD in three dimensions (3D-EDD) as plotted in fig. 5 and in two dimensions (2D-EDD) for the plane (110) as plotted in fig. 6. Each plot reveals that one set of EDD is the opposite of the other along the c axis. There are regions of charge build-up corresponding of bonding regions that are paired up with regions of charge depletion corresponding to anti-bonding regions along the c axis. One notes that the regions of charge build-up and charge depletion will oscillate with the $E_{2g}$ frequency mode in case of $MgB_2$. The electron-hole pairing obtained is in a real space characterized by Charge and Parity symmetries, or the so-called CP symmetry. One notes also that the increase of the amount of charge density transferred from the bonding to the anti-bonding regions as the separation along the c axis between the atoms Mg-B is decreased. This distance is temperature dependent and modulated by the two vibrations modes $A_{2u}$ and $E_{2g}$ in case of $MgB_2$.

## 3. The case of Hydrides or Hydrogen based materials

$H_3S$ is obtained from $H_2S$ under ultrahigh pressure [19] and superconducts at high $T_c$ ~203 K. Its crystal structure is a bcc structure ( $Im\bar{3}m$, a = 3.089 Å). First principle calculation of EDD is determined by subtracting the electron density of the crystal distorted by frozen-phonon modes at the Brillouin zone boundary from that of the perfect crystal. The phonon mode used for calculation is depicted in fig.7 a. The EDD calculated, plotted in fig 7. b and fig. 8 correspond to 3D-EDD plot and 2D-EDD plot of the plane (010) respectively. These plots reveal a regions of charge build-up corresponding of bonding regions that are paired up with regions of charge depletion corresponding to anti-bonding regions along the c axis where the symmetry obtained is also in this case a CP symmetry.

Another compound of the family of hydrides is Lanthanum decahydride $LaH_{10}$ which is also obtained under ultrahigh pressure and superconducts at very high critical temperature 250 K [20, 21]. Its crystal structure is cubic $Fm\bar{3}m$, a= 4.3646 Å at 152 GPa. First principle calculation of the EDD is also performed by subtracting the electron density of the crystal distorted by frozen-phonon modes at the Brillouin zone from that of the perfect crystal. The EDD calculated, plotted in figs. 9 and 10 correspond to 3D-EDD plot and 2D-EDD plot of the plane (100) respectively.

These plots reveal a regions of charge build-up corresponding of bonding regions that are paired up with regions of charge depletion corresponding to anti-bonding regions along the c axis where the symmetry obtained is always a CP symmetry.

## 4. The case of simple metals of the periodic table

A series of EDD calculation is performed for simple metals of the periodic table; Lead Pb ($T_c$= 7.193 K), Vanadium V ($T_c$= 5.38 K), Aluminum Al ($T_c$= 1.180 K), Copper Cu (not superconducting) and Iron Fe (not superconducting at normal pressure) [22].

3D-EDD for the three superconducting elements Pb, V and Al are illustrated in figs 11 a, b and c respectively, which all reveal a regions of charge build-up corresponding of bonding regions paired up with regions of charge depletion corresponding to anti-bonding regions along the c axis. The symmetry obtained is also in this case a CP symmetry. These results can be well understood in terms of phonon-distortion driven local polarizability or change of orbital hybridization as evidenced by 2D plot of the electron charge density of Pb along (100) plane depicted in fig. 12 a and b for the perfect crystal and that of the distorted one respectively.

However, in the case of copper Cu which is known to not superconduct at any given temperature, 3D-EDD does not reveal any regions of charge build-up corresponding of bonding regions nor regions charge depletion corresponding to anti-bonding regions as it is the case of Pb, Al and V (see fig. 13). This result can be well understood in terms of phonon-distortion driven no polarizability of copper atoms as evidenced by 2D plot of the charge density of Cu along (100) plane of fig. 14.

For Iron Fe, 3D-EDD illustrated in fig. 15, does reveal a region of charge build-up corresponding of bonding regions and charge depletion corresponding to anti-bonding regions as in case of Pb, Al and V so it is expected that this element will superconduct, however it does not superconduct at any given temperature. The reason is that this element shows a magnetic order that breaks the Time reversal symmetry and since the combined CP and Time reversal symmetry the so-called CPT is a preserved symmetry of nature, CP symmetry is also broken in that case and no superconductivity would manifest as consequence of this CP symmetry broken.

## 5. Conclusion

Electron density dynamic's response to longitudinal acoustical phonon modes at the Brillouin zone boundary is proven to be the key factor in the superconducting mechanism in $MgB_2$ compound as well as in all superconducting compounds leading to a real space electron-hole pairing characterized by CP symmetry (Charge and Parity symmetries) that enables a unified treatment of superconductivity phenomenon. First principle calculation of the EDD has been determined by subtracting the electron density of the crystal distorted by longitudinal acoustical frozen-phonon mode at the Brillouin zone boundary from that of the perfect crystal. The origin of this electron density response to this kind of phonon-distortion that leads to a real space electron-hole pairing characterized by CP symmetry, has been discussed in terms of local polarizabilities and delocalized transfer of charge caused by phono-distortion driven dynamical change of orbital hybridizations.

This approach also provides clear and quick guidelines in the search for new superconductors.

**Acknowledgment**

The author would like to thank Dr S. Toumi for the correction of the present manuscript and for numerous fruitful discussions.

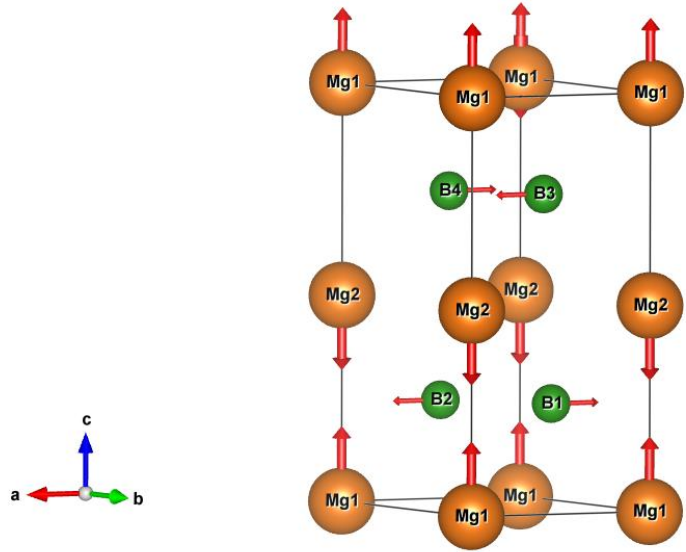

**Figure 1.** Supercell of the Hexagonal crystal structure of $MgB_2$ ($AlB_2$-type structure with a space group of P6/mmm). Arrows denote the direction of atomic displacement patterns for $E_{2g} + A_{2u}$ modes at the zone-boundary A point of the hexagonal Brillouin zone. While $E_{2g}$ is an optical mode that involves in-plane displacements of B atoms, the acoustical $A_{2u}$ vibration mode involves the displacement of Mg atoms along the c axis. Note also that a compressed and elongated cell will alternate along the c axis.

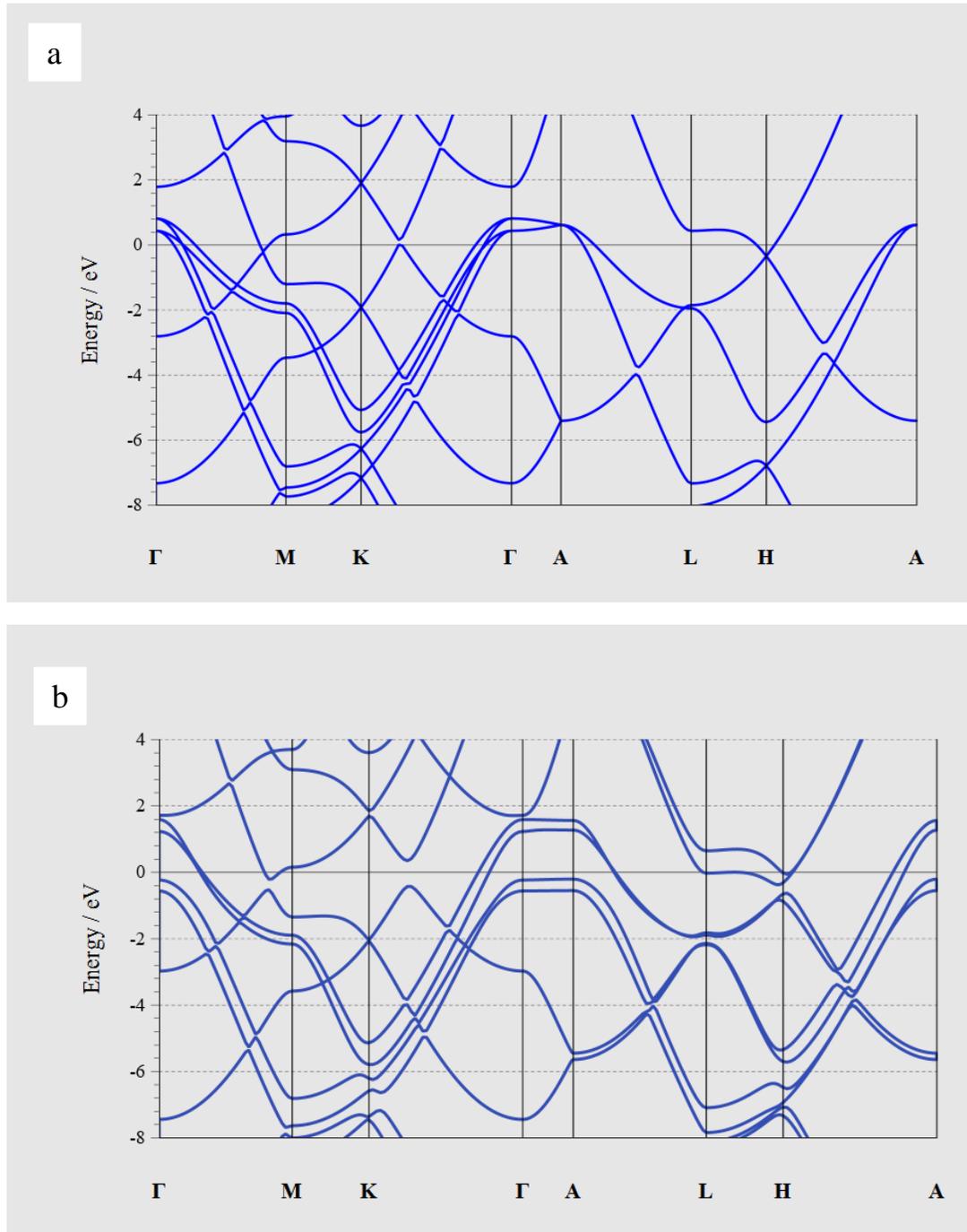

**Figure 2.** Electronic band structure calculated for a) the perfect supercell of MgB$_2$, b) the phonon-frozen distorted supercell. While the $\sigma$ bands of the compressed cell move down the Fermi level indicating almost transition to insulating state, $\sigma$ bands of the elongated cell move upward indicating an increase of the hole doping level. One notes the oscillation of the metallic-insulating state transition with the A$_{2u}$ frequency.

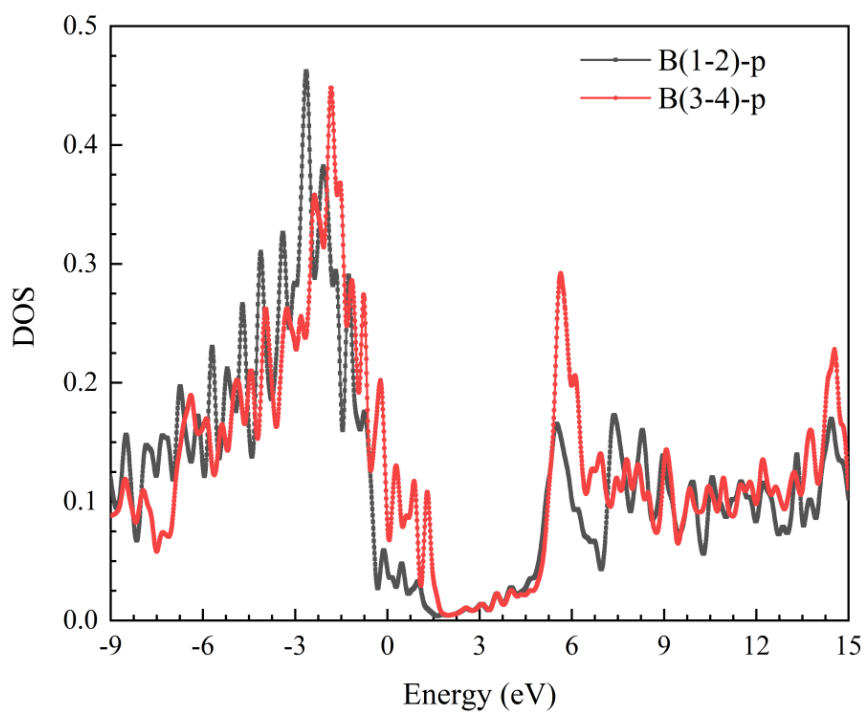

**Figure 3.** Evidence of the insulating state transition of the compressed cell from partial density of *p* electronic states (PDOS) which reveals the shift in energy of B (1) and B (2) electronic states down the Fermi level.

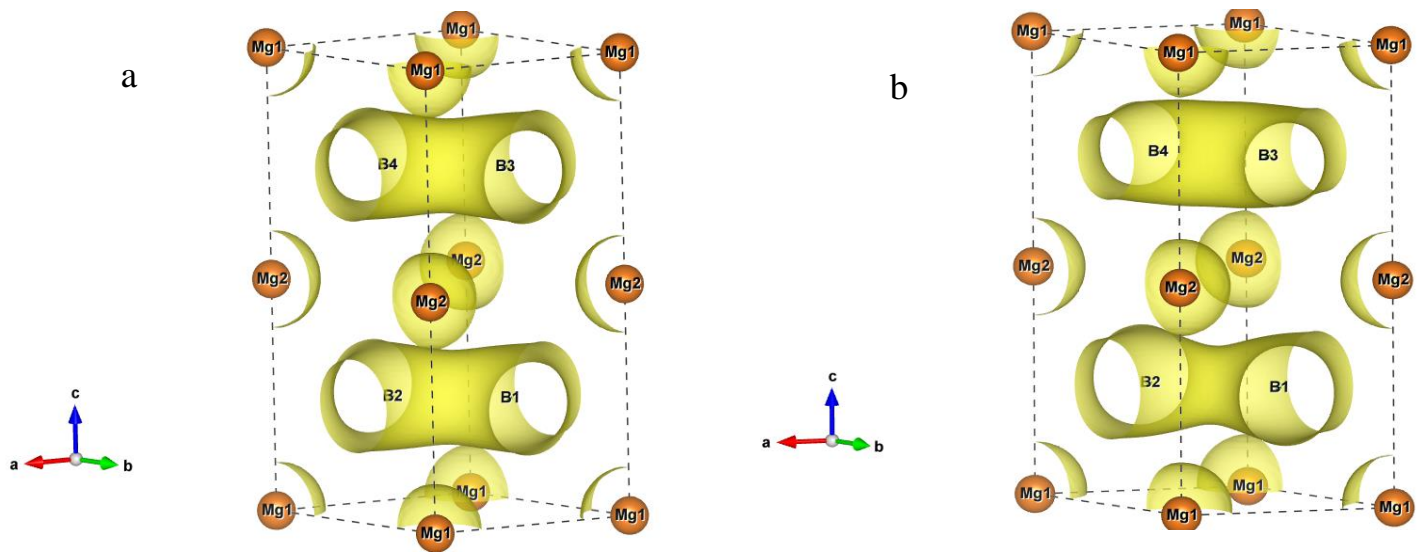

**Figure 4.** Electron density shown with an isosurface of 0.5 Å$^{-3}$ calculated for a) the perfect supercell of MgB$_2$, b) the phonon-frozen distorted supercell. A comparison between them reveals that there is more electronic polarizability in the compressed cell and less polarizability in the elongated cell relative to that of the perfect crystal.

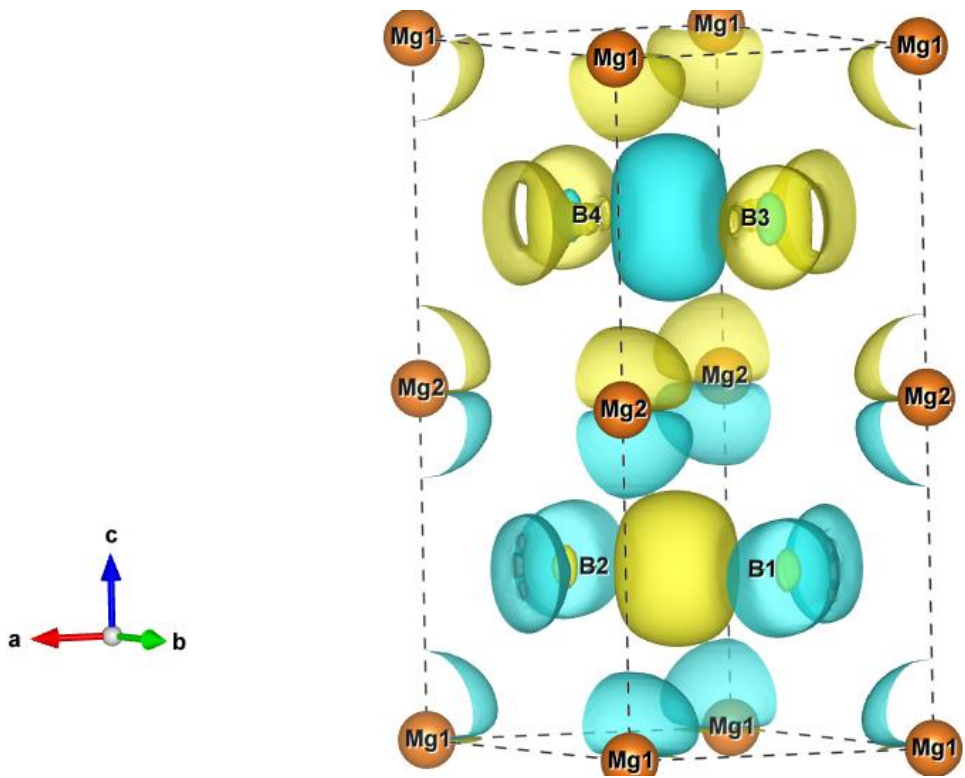

**Figure 5.** Electron density difference 3D-EDD, shown with an isosurface level of 0.05 Å$^{-3}$, calculated by subtracting the electron density of the phonon-distorted crystal from that of the perfect crystal of MgB$_2$ supercell. The yellow and the blue surfaces denote positive and negative densities respectively. The regions of charge build-up are paired up with regions of charge depletion along the c axis. These regions of charge build-up and charge depletion will oscillate with the E$_{2g}$ frequency mode in case of MgB$_2$.

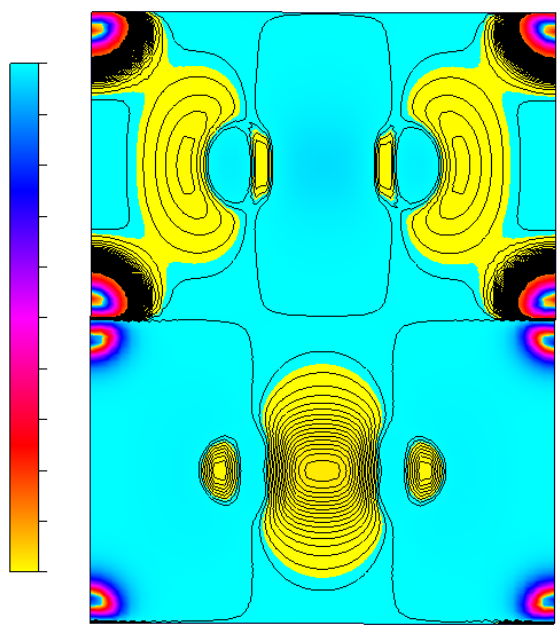

**Figure 6.** Electron density difference distribution 2D-EDD on the plane (110) of the supercell. Contours are plotted with an interval of 0.01 Å$^{-3}$. The yellow and the green surfaces denote positive and negative densities respectively. The regions of bonding paired up with regions of antibonding will oscillate with the E$_{2g}$ frequency mode.

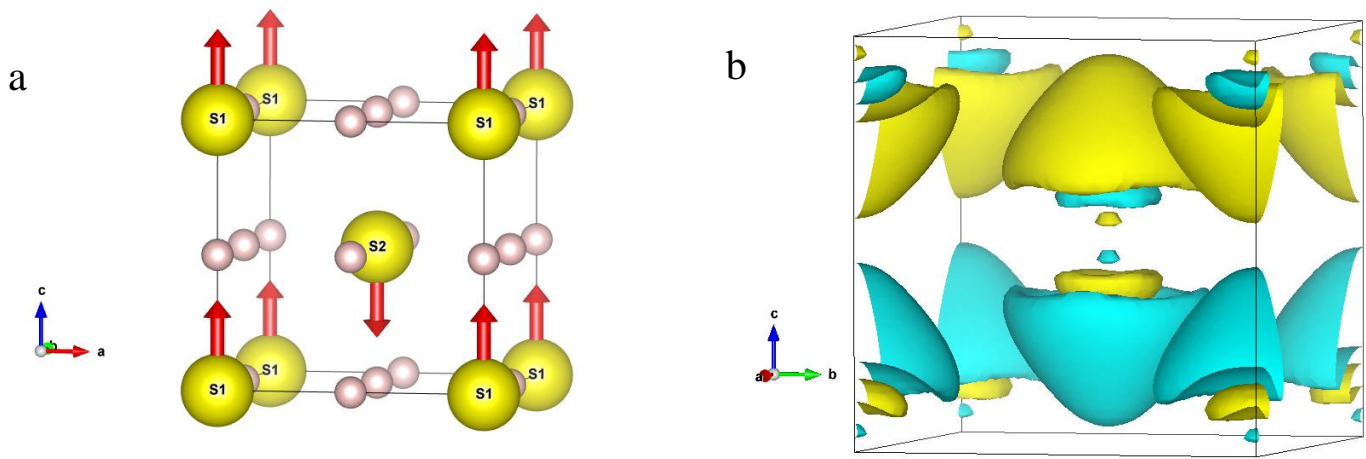

**Figure 7.** a) crystal structure of H$_3$S along with frozen phonon at the Brillouin zone boundary along the c axis. Arrows denote displacements of atoms. b) Electron density difference 3D-EDD shown with an isosurface level of 0.03 Å$^{-3}$ for H$_3$S superconductor. The lattice constant used is that of room temperature.

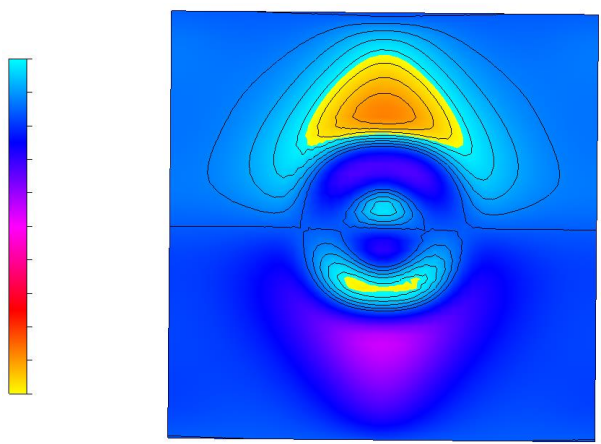

**Figure 8.** 2D-EDD isocontours along the plane (100) plotted with an interval of 0.01 Å$^{-3}$ for H$_3$S crystal. The bonding regions are paired up with the antibonding regions along the c axis (for phonon-distortion parallel to z).

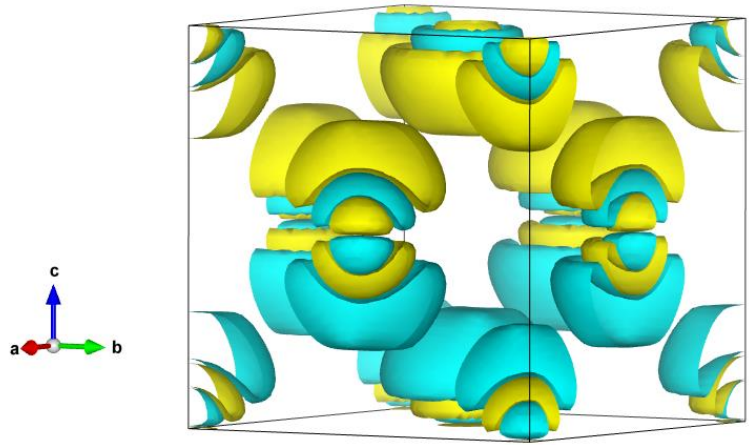

**Figure 9.** Electron density difference 3D-EDD of LaH$_{10}$ shown with an isosurface level of 0.05 Å$^{-3}$. The lattice constant used is that of room temperature.

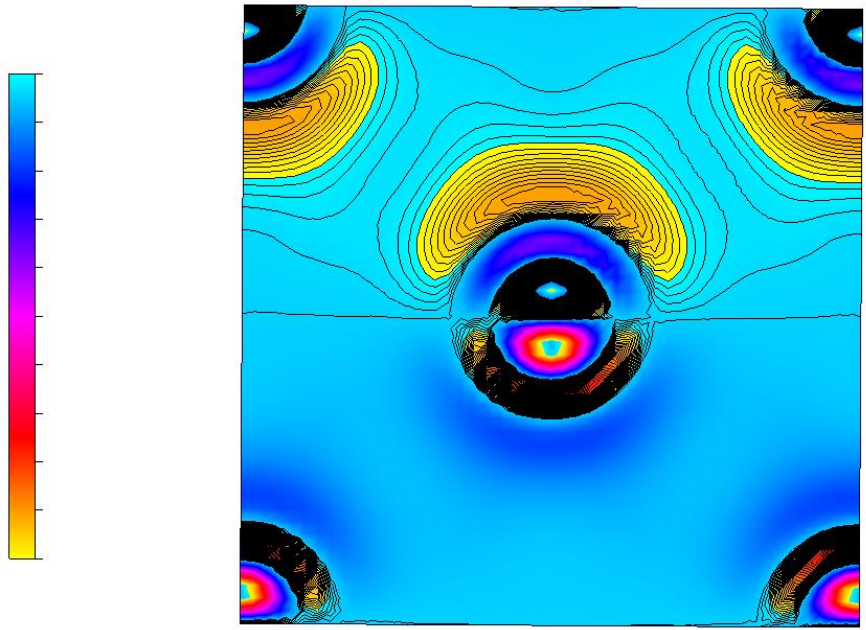

**Figure 10.** 2D-EDD isocontours along the plane (100) of LaH$_{10}$ crystal plotted with an interval of 0.01 Å$^{-3}$. The bonding regions are paired up with the antibonding regions along the c axis (for phonon-distortion parallel to z).

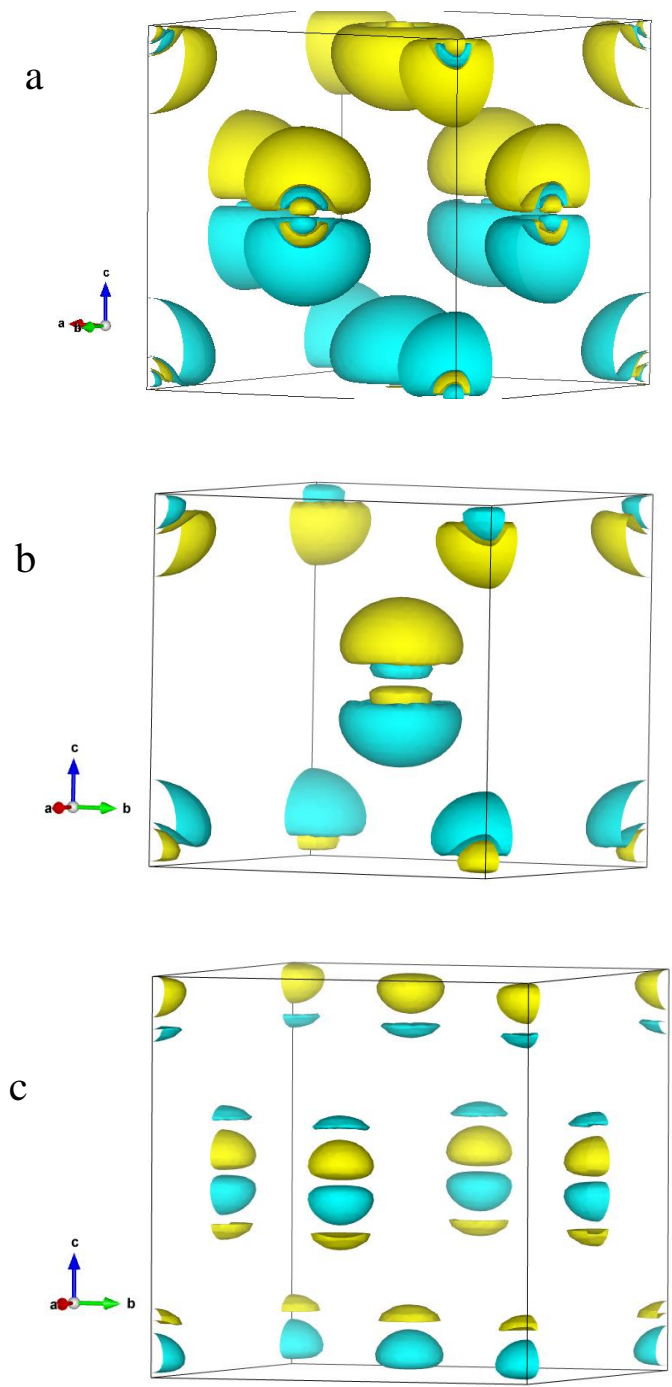

**Figure 11.** 3D-EDD isosurfaces of a) Lead Pb (isosurface level of 0.05 Å$^{-3}$,) b) Vanadium V (isosurface level of 0.3 Å$^{-3}$), c) Aluminium Al (isosurface level of 0.015 Å$^{-3}$). The lattice constants used are those of room temperature.

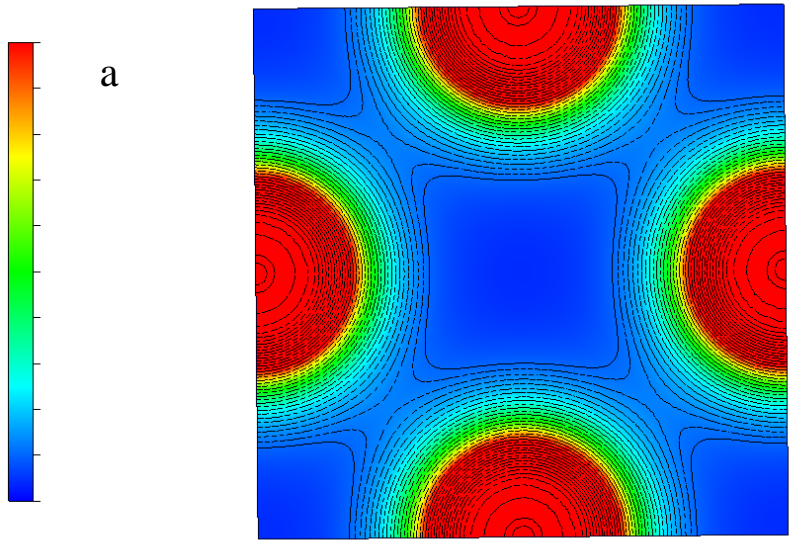

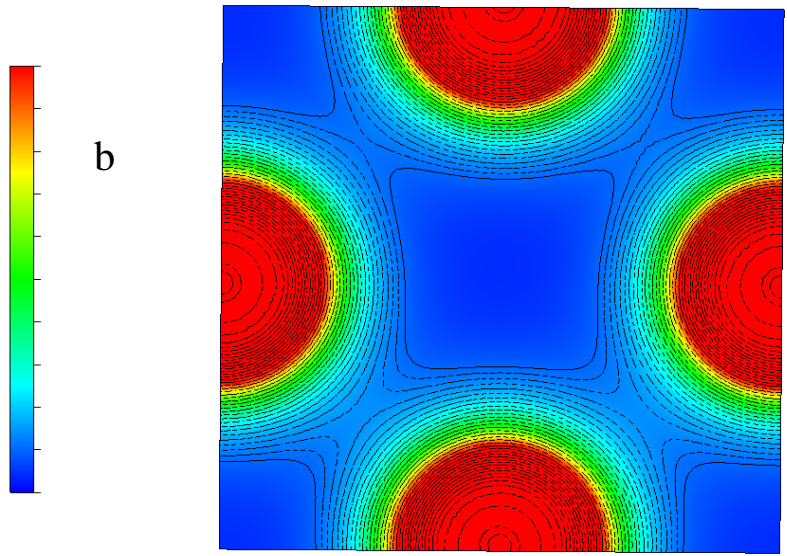

**Figure 12.** Logarithmic 2D electron charge density isocontours for the crystal of Lead (Pb) along the plane (100). a) for a perfect crystal, b) for a distorted crystal with a frozen phonon at the Brillouin zone along the c axis.

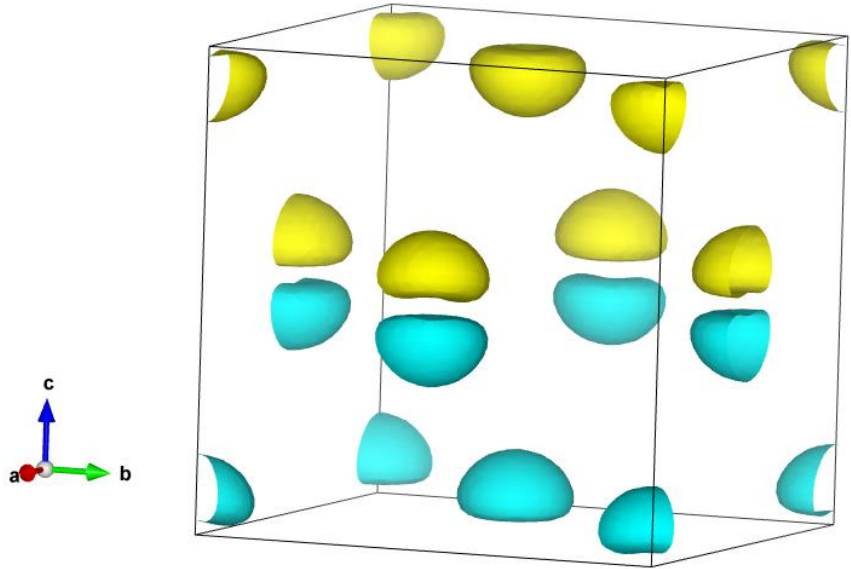

**Figure 13.** 3D-EDD isosurfaces shown with an isosurface level of 0.7 Å$^{-3}$ of Copper (Cu) which is known to not superconduct. Note the absence of the bonding regions that are paired up with the antibonding regions along the c axis in case of superconducting materials

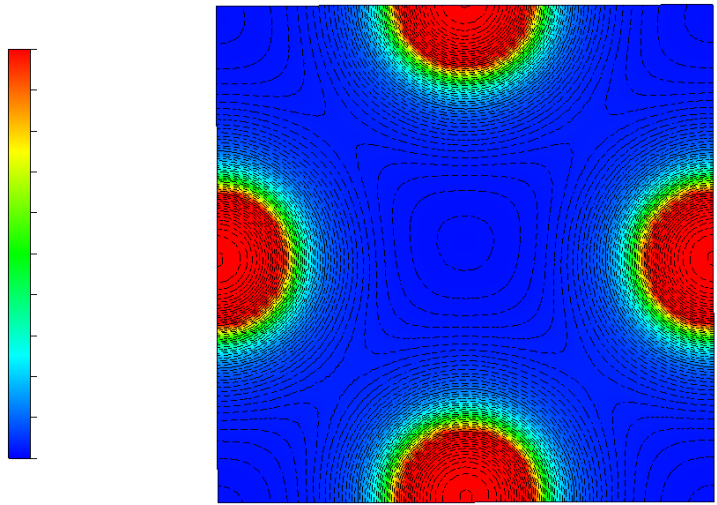

**Figure 14.** Logarithmic 2D plot in (100) plane of the electron charge density of Cu distorted with frozen phonon at the Brillouin zone boundary along the c axis. Note the absence of change in orbital hybridizations along the c axis. (Compare with fig.12 b)

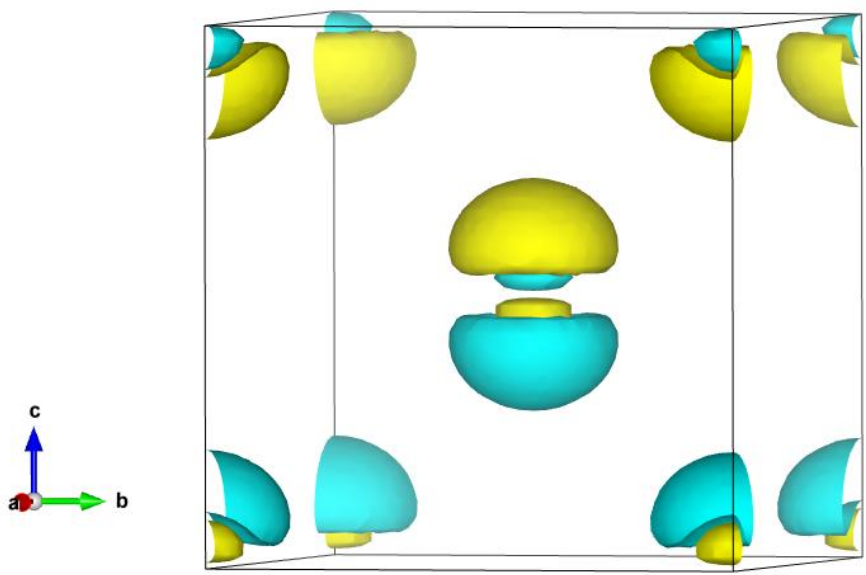

**Figure 15.** 3D-EDD isosurfaces of Iron Fe shown with an isosurface level of 0.7 Å$^{-3}$. Although the bonding regions that are paired up with the antibonding regions along the c axis are present as in case of superconducting materials, this material does not superconduct because of the magnetic order that breaks the Time reversal symmetry.